 \documentclass[review,3p,authoryear]{elsarticle}
\usepackage{graphicx}
\usepackage[utopia]{mathdesign}
\usepackage[OMLmathrm,OMLmathbf]{isomath} % options define which alphabets will be loaded, i.e. if bold face font is not necessary, `OMLmathbf` can be ommitted.
    \usepackage{lmodern}
\usepackage{amsmath}
\usepackage{graphicx}
\usepackage{dsfont}
\usepackage{color}
\usepackage{tensor}
\usepackage[mathscr]{eucal}
\numberwithin{equation}{section}

\def \dex{{\mathrm{d}\,}}
\def \n{{\mathbfit{n}}}
\def \m{{\mathbfit{m}}}

\def \b{{\mathbfit{b}}}

\def \u{{\mathbfit{u}}}
\def \e{{\mathbfit{e}}}

\def \q{{\mathbfit{q}}}
\def \x{{\mathbfit{x}}}

\def \rhobf{{\mathbfit{\rho}}}

\def \Psibf{{\mathbfit{\Psi}}}

\def \sigmabf{{\mathbfit{\sigma}}}
\def \betabf{{\mathbfit{\beta}}}

\def \updelta{{\mathrm{\delta}}}
\def \upvarepsilon{{\mathrm{\varepsilon}}}
\def \sign{{\operatorname{sign}}}
\def \dex{{\operatorname{d}}}
\def \partialvar{{\partial}}
\journal{Journal of the Mechanics and Physics of Solids}

\begin{document}

\begin{frontmatter}

%% Title, authors and addresses

%% use the tnoteref command within \title for footnotes;
%% use the tnotetext command for theassociated footnote;
%% use the fnref command within \author or \address for footnotes;
%% use the fntext command for theassociated footnote;
%% use the corref command within \author for corresponding author footnotes;
%% use the cortext command for theassociated footnote;
%% use the ead command for the email address,
%% and the form \ead[url] for the home page:
%% \title{Title\tnoteref{label1}}
%% \tnotetext[label1]{}
%% \author{Name\corref{cor1}\fnref{label2}}
%% \ead{email address}
%% \ead[url]{home page}
%% \fntext[label2]{}
%% \cortext[cor1]{}
%% \address{Address\fnref{label3}}
%% \fntext[label3]{}
%% \listofchanges
\title{Thermodynamically consistent continuum dislocation dynamics}

%% use optional labels to link authors explicitly to addresses:
%% \author[label1,label2]{}
%% \address[label1]{}
%% \address[label2]{}
\author[label1]{Thomas Hochrainer}

\ead{hochrain@uni-bremen.de}
\address[label1]{BIME - Bremer Institut f\"ur Strukturmechanik und Produktionsanlagen, Universit\"at Bremen, IW3, Am Biologischen Garten 2, 28359 Bremen, Germany}

\begin{abstract}
Dislocation based modeling of plasticity is one of the central challenges at the crossover of materials science and continuum mechanics. Developing a continuum theory of dislocations requires the solution of two long standing problems: (\emph{i}) to represent dislocation kinematics in terms of a reasonable number of variables and (\emph{ii}) to derive averaged descriptions of the dislocation dynamics (i.e. material laws) in terms of these variables. The kinematic problem (\emph{i}) was recently solved through the introduction of continuum dislocation dynamics (CDD), which provides kinematically consistent evolution equations of dislocation alignment tensors, presuming a given average dislocation velocity (Hochrainer (2015), Philos. Mag. 95 (12), 1321--1367). In the current paper we demonstrate how a free energy formulation may be used to solve the dynamic closure problem (\emph{ii}) in CDD. We do so exemplarily for the lowest order CDD variant for curved dislocations in a single slip situation. In this case, a thermodynamically consistent average dislocation velocity is found to comprise five mesoscopic shear stress contributions. For a postulated free energy expression we identify among these stress contributions a back-stress term and a line-tension term, both of which have already been postulated for CDD. A new stress contribution occurs which is missing in earlier CDD models including the statistical continuum theory of straight parallel edge dislocations (Groma et al. (2003), Acta Mater. 51, 1271-1281). Furthermore, two entirely new stress contributions arise from the curvature of dislocations.
\end{abstract}

\begin{keyword} Crystal plasticity \sep
Continuum theory of dislocations \sep Thermodynamic consistency

%% keywords here, in the form: keyword \sep keyword

%% PACS codes here, in the form: \PACS code \sep code

%% MSC codes here, in the form: \MSC code \sep code
%% or \MSC[2008] code \sep code (2000 is the default)

\end{keyword}

\end{frontmatter}

\section{Introduction}

Plastic deformation of crystals is driven and controlled by the motion and interaction of dislocations. Dislocations are line like crystal defects which move when subjected to shear stresses and leave a permanent shear of the crystal when moving. Typically, metals contain a larger number of dislocations with line densities $\rho$ in the range of $10^{10}$ to $10^{14} \, \text{m}^{-2} $. These high numbers suggest that it should be possible to formulate a statistical mechanics theory of dislocations as a basis for crystal plasticity. However, as yet there is only a statistical theory available for strongly simplified systems of straight parallel edge dislocations \citep{groma_cz03}. This theory was originally developed from a Bogoliubov-Born-Green-Kirkwood-Yvon hierarchy adopted from interacting particle systems. For this derivation pair correlation functions have been obtained from statistical evaluation of a large number of quasi two-dimensional (2D) discrete dislocation simulations. Subsequently, part of the theory was recast into a phase-field type thermodynamic description \citep{groma_gk07} derived from an effective free energy. A similar free energy for systems of straight parallel edge dislocations has recently been derived from a partition function by \citet{kooiman_hg15}. Transferring the statistical mechanics or thermodynamically inspired methods to systems of curved dislocations was doomed to failure as long as no suitable density variables capable of reflecting the kinematics of moving flexible lines were available. Such kinematically sound density variables have only recently been presented with the so called continuum dislocation dynamics (CDD) framework by \citet{hochrainer_etal14} and \citet{hochrainer15}. A transfer of the statistical methods to curved dislocations seems impractical because of the difficulty to obtain sufficient statistics from three-dimensional discrete dislocation simulations. However, in the following we show that a thermodynamic approach is adoptable for curved dislocations.

In the current paper we develop a thermodynamic formulation exemplarily for the lowest order CDD variant for curved dislocations in a single slip situation. The main result is that in this case a thermodynamically consistent average dislocation velocity is found to comprise five mesoscopic shear stress contributions as opposed to only one mesoscopic (back-) stress found in the quasi 2D case so far. For a postulated free energy expression we identify among these stress contributions a generalization of the quasi 2D back-stress term and a line-tension term, which have both been postulated in similar form for CDD before. Notably, a new stress contribution occurs which is missing in earlier CDD models including the quasi 2D case. Furthermore, two entirely new mesoscopic stress contributions are found to arise from the curved nature of dislocations.

%In the current paper we first derive a thermodynamically consistent dislocation velocity for CDD without specifying the free energy. Subsequently, we define a phenomenological free energy functional, the form of which is assumed in analogy to the available theory for quasi two-dimensional systems. From requiring thermodynamic consistency with this free energy we obtain five mesoscopic stress contributions to the dislocation velocity which generalize terms known from the quasi 2D theory. Furthermore, we not only find new terms unique to the system of curved dislocations, but also a term, the specialization of which is missing in the original quasi 2D theory.

The paper is structured as follows: in Section \ref{sec: notations} we briefly introduce notations and preliminaries from elasticity and crystal plasticity in small deformations. The quasi 2D theory of straight parallel edge dislocations serves both as point of departure and as reference in the sequel. We therefore recall the quasi 2D theory as derived from statistical mechanics and its thermodynamic reformulation in Section \ref{sec: quasi 2d}. In Section \ref{sec: curved CDD} we briefly introduce the kinematics of the lowest order CDD for curved dislocations before we derive the general form of a thermodynamically consistent dislocation velocity $ v $ within this framework. Subsequently, we assume a free energy expression for the curved dislocation case in analogy to the one used in quasi 2D CDD. This free energy is used to illustrate the meaning of five mesoscopic stress contributions to the thermodynamically consistent dislocation velocity and to relate these stress contributions to expressions obtained in quasi 2D CDD and postulated for CDD of curved dislocations before. In Section 5 we summarize and discuss the results and depict open problems.
\section{Notations and preliminaries} \label{sec: notations}
We employ the convention that scalars are denoted with light face symbols while vectors and tensors are denoted with boldface symbols. Note, however, that we do not employ any notational difference between tensors of different orders, including vectors. Frequent use of coordinate notation shall avoid any ambiguities in this regard. We exclusively work in Cartesian coordinates with unit vectors $ \left\{ \e_i; \; i=1,2,3 \right\} $. Accordingly, we do not distinguish co- and contra-variant indices and employ the modified Einstein summation convention, where summation applies to all pairs of equal (lower) indices. The Kronecker delta (metric tensor) is denoted with $ \updelta_{ij} $ and the totally antisymmetric Levi-Civita-symbol reads $\upvarepsilon_{ijk}$.
 
Partial derivatives in the coordinate directions shall be denoted with $\partial_i$; the partial time derivative is usually denoted with $ \partial_t $. We do not apply an according short hand notation for derivatives of functional kernels with respect to their arguments.
\subsection{Kinematics}\label{kinematics small strain}
We restrict attention to small deformations such that the deformation is described by the displacement vector field $ \u $ and the distortions by the displacement gradient $ \nabla \u $. In compliance with the small strain assumption we assume an additive split of the total distortion
\begin{equation}
	\nabla \u  = \mathbfit{\beta}^\mathrm{el} + \mathbfit{\beta}^\mathrm{pl},
\label{eq: small strain}
\end{equation}
into elastic and plastic distortion tensors, respectively.

In crystal plasticity, plastic slip is confined to well defined slip planes which we characterize by slip plane normal $ \n $ and slip direction $ \m $. The plastic distortion tensor in small deformations is additively composed of slip system specific shear tensors. In the sequel \emph{we only consider one active slip system}. Therefore, the plastic distortion tensor is of the form
\begin{equation}
	\mathbfit{\beta}^\mathrm{pl} = \gamma \n \otimes \m,
\label{eq: beatpl}
\end{equation}
with $ \gamma $ denoting the accumulated plastic slip on the slip system. We note that the slip direction $ \m$ is the normalized Burgers vector $\b$ of dislocations on the slip system, i.e. $ \m  = \b / b $, where $ b  $ denotes the modulus of the Burgers vector.

The total strain tensor $ \mathbfit{\epsilon} $ derives from symmetrizing the displacement gradient
\begin{equation}
	\epsilon_{ij} = \frac{1}{2 }\left( \partial_j u_i + \partial_i u_j \right).
\label{eq: epsilon small strain}
\end{equation}
Accordingly, also the total strain is additively split into an elastic and a plastic strain, $\mathbfit{\epsilon}  = \mathbfit{\epsilon}^\mathrm{el} + \mathbfit{\epsilon}^\mathrm{pl}$, which each time arise from symmetrizing the respective distortion tensors.

\subsection{Constitutive theory}
We assume linear elasticity such that the elastic constitutive equation connecting the stress tensor $ \sigmabf $ with strain is
\begin{equation}
	\sigma_{ij} = C_{ijkl} \left( \partial_k u_l - \beta^\mathrm{pl}_{kl} \right) = C_{ijkl} \epsilon^\mathrm{el}_{kl}  .
\label{eq: linear elasticity}
\end{equation}
Volumetric forces will not be considered in this paper. Hence, conservation of momentum yields 
\begin{equation}
	\partial_j \sigma_{ji} = 0.
\label{eq: divergence stress}
\end{equation}
Due to conservation of moment of momentum the stress tensor is symmetric, $ \sigma_{ij} = \sigma_{ji} $.

In the sequel we employ isotropic elasticity, such that we may write the elasticity tensor in terms of bulk modulus $K$ and shear modulus $G$ as
\begin{equation}
	C_{ijkl} = K \updelta_{ij} \updelta_{kl} + G \left( \updelta_{ik}\updelta_{jl} + \updelta_{il}\updelta_{jk} - \frac{2}{3} \updelta_{ij} \updelta_{kl} \right).
\label{eq: isotropic linear elasticity}
\end{equation}
Note that the restriction to isotropic elasticity is convenient but not essential for the results of this paper.

\subsection{Boundary conditions and partial integration}
In the current paper we concentrate on bulk behavior and regard an infinite crystal which is loaded by a remotely applied external stress $\mathbfit{\sigma}^\mathrm{ex}$. We frequently use partial integration, where we always assume that boundary terms vanish, e.g. by assuming that the involved variables vanish in the infinite or by employing periodic boundary conditions. In other words, we use
\begin{equation}
	\int \partial _i f \; g \; \dex V = - \int f \; \partial_i g \;\dex V
\label{eq: partial integration}
\end{equation}
for suitable functions $f$ and $g$.
\section{Continuum dislocation dynamics of straight parallel edge dislocations} \label{sec: quasi 2d}
The statistical mechanics theory for systems of straight parallel edge dislocations developed by \citet{groma_cz03} is as yet the only statistical continuum theory of dislocations derived from systematic averaging. The theory was successfully shown to produce essentially equivalent results to discrete dislocation simulations of straight edge dislocations treated as point particles with two possible signs in an elastic medium \citep{yefimov_gg04}. The theory henceforth serves as point of departure and as reference in the current paper.

In the case of only edge dislocations we choose the coordinate system such that the Burgers vector points in 1-direction, $ \b = b \e_1 $, and positive edge dislocations point in 2-direction. The slip plane normal accordingly points in 3-direction, $ \n = \e_3 $. Assuming additionally a plane strain state this dislocation system and its evolution can be completely described as quasi two-dimensional system in the 1-3 plane. 
\subsection{Continuum kinematics of straight parallel edge dislocations}
The dislocation system is described in terms of the so-called total dislocation density $ \rho $ and the net-dislocation density or geometrically necessary dislocation (GND) density $ \kappa $. The kinematics of plastic deformations as well as the evolution of the dislocation variables are described in terms of an average dislocation velocity $ v$. For the regarded single slip system the plastic distortion tensor has the form $ \betabf^\mathrm{pl} = \gamma \e_3 \otimes \e_1$, such that $ \gamma = \beta^\mathrm{pl}_{31} $. The evolution of the plastic slip is obtained from Orowan's equation. The evolution of the plastic variables consequently read (cf. \citet{groma_cz03})
\begin{eqnarray} 
%	\partial_t \beta^\mathrm{pl}_{21}  &=& \partial_t \gamma = \rho v b \label{Eq: evol gamma} \\
	\partial_t \gamma &=& \rho v b, \label{Eq: evol gamma} \\
	\partial_t \rho  &=&  -\partial_1 \left( v \kappa \right), \label{Eq: evol rho istvan} \\
	\partial_t \kappa	&=& -\partial_1 \left( v \rho \right).  \label{Eq: evol kappa istvan}
\end{eqnarray}
Before moving on to the constitutive equations we note that from the above one easily realizes the relation $ \kappa = - \partial_1 \gamma /b $. We conclude that $\kappa$ defines the only non-vanishing component $ \alpha_{21} = \kappa b $ of the dislocation density tensor $ \alpha_{ij} = \upvarepsilon_{ikl} \partial_k \beta^\mathrm{pl}_{lj} $ \citep{kroener58}. This justifies the denomination of $ \kappa $ as the GND density.
\subsection{Constitutive theory for straight dislocations} \label{sec: constitutive straight}
\subsubsection{Results from statistical mechanics approach}

In order to employ the kinematic evolution equations (\ref{Eq: evol gamma})--(\ref{Eq: evol kappa istvan}) as a material law, a constitutive closure is needed which defines the average dislocation velocity $ v $ in terms of the current stress and dislocation state. The dislocation velocity was obtained from statistical averaging in the quasi 2D theory of straight parallel edge dislocations \citep{groma_cz03, yefimov_gg04} in the form
\begin{equation}
	 v = M b \, \sign \left({\tau_\mathrm{net}}\right) \left\langle \left| \tau_\mathrm{net} \right|  -  \tau_\mathrm{f} \right\rangle ,
\label{eq: velocity assumption}
\end{equation}
where $M$ is a dislocation mobility coefficient, the function \emph{sign} returns the signature of its argument and $\left\langle \cdot \right\rangle$ denote the Macaulay brackets, which return their argument if it is positive and zero else. The so-called flow-stress $\tau_\mathrm{f}$ defines a threshold (shear) stress such that the velocity is zero as long as the absolute value of the net-shear stress $ \tau_\mathrm{net} $ is smaller than the threshold. The flow stress is well known to obey the form $ \tau_\mathrm{f} = \alpha G b \sqrt{\rho}$ (called the Taylor relation), with a dimensionless coefficient $\alpha $ typically found between 0.2 and 0.5 (cf., e.g. \citet{mecking_k81}). This form for $ \tau_\mathrm{f} $ could also be derived from rigorous averaging in the quasi 2D case \citep{groma_cz03}. The net shear stress was derived in the same work as 
\begin{equation}
	  \tau_\mathrm{net} :=  \tau_\mathrm{ex} + \tau_\mathrm{sc} - \tau_\mathrm{b}.
\label{eq: v stress groma}
\end{equation}
The appearing stress contributions are the external shear stress $ \tau_\mathrm{ex} = \sigma^\mathrm{ex}_{ij} n_j m_i $ from remote boundary conditions independent of dislocations and the self-consistent stress $ \tau_\mathrm{sc} $, which accounts for the long range stress field due to distributions of geometrically necessary dislocations. The self-consistent stress is obtained as
\begin{equation}
	  \tau_\mathrm{sc} \left( \x \right):=  \int{\kappa\left( \x' \right) \tau_\mathrm{ind} \left( \x, \x' \right)} \dex V',
\label{eq: tausc}
\end{equation}
where $ \tau_\mathrm{ind} \left( \x, \x' \right) $ denotes the shear stress induced at location $ \x $ by a dislocation at position $ \x' $ (see, e.g. \citet{hirth_l68}). Furthermore, there appears a mesoscopic stress contribution called the back-stress, which is of the form
\begin{equation}
	  \tau_\mathrm{b} = \frac{B Gb}{\rho} \partial_1 \kappa = - \frac{B G}{\rho} \partial_1 \partial_1 \gamma,
\label{eq: v stress assumption}
\end{equation}
with a dimensionless constant  $ B $. Note that we use the sign convention for the back stress employed in \citet{yefimov_gg04}, which differs from the one in \citet{groma_cz03}.

In the sequel we will abandon the distinction between the external stress and the self-consistent stress as they sum up to the macroscopic or mean field resolved shear tress $ \tau = \tau_\mathrm{ex} + \tau_\mathrm{sc}$. The mean field shear stress is simply the resolved shear stress from the stress tensor $\mathbfit{\sigma}$ which is obtained upon solving for the displacement field $ \u $ in (\ref{eq: linear elasticity}) subject to the solenoidality of the stress tensor, given the accumulated plastic slip $ \gamma $ and remote boundary conditions $\sigmabf^\mathrm{ex}$, cf., e.g. \citet{elazab00, yefimov_gg04, sandfeld_mz13}. In other words, we find $ \tau = \sigma_{21} $ and rewrite (\ref{eq: v stress groma}) as
\begin{equation}
	  \tau_\mathrm{net} := \tau - \tau_\mathrm{b} .
\label{eq: v stress groma 2}
\end{equation}
We note that the use of the eigenstrain formulation instead of the convolution (\ref{eq: tausc}) for the self-consistent stress naturally allows for dropping the assumption of isotropic elasticity.
\subsubsection{Phase field type reformulation}
The net shear stress $ \tau_\mathrm{net} $ introduced above plays the role of a driving force for the evolution of the dislocation system. Originally derived from statistical averaging, it was later obtained in a phase field type description as a variational derivative of a free energy of the dislocation system. However, the flow stress $ \tau_\mathrm{f} $ is not derivable in this way. It was therefore suggested by \citet{groma_vi15} that the flow stress should be considered as a friction-type ingredient of a non-constant mobility function. With such a mobility function we assume the dislocation velocity to have the generic form  
\begin{equation}
	  v = M(\tau_\mathrm{net},\rho) b \tau_\mathrm{net}.
\label{eq: v with mobility}
\end{equation}
In order for this form to be consistent with (\ref{eq: velocity assumption}) the mobility function must be defined as
\begin{equation}
	  M(\tau_\mathrm{net},\rho) = M \frac{ \left\langle \left| \tau_\mathrm{net} \right|  -  \tau_\mathrm{f}(\rho) \right\rangle }{\left| \tau_\mathrm{net}\right|}.
\label{eq: mobility}
\end{equation}
Note that the mobility vanishes in the case of a vanishing net shear stress as it vanishes whenever $  \left| \tau_\mathrm{net} \right| \leq \tau_\mathrm{f}$.

The free energy $ \Psibf = \Psibf^\mathrm{el} +\Psibf^\mathrm{def} $ of the quasi 2D CDD system is composed of a `classical' elastic energy $ \Psibf^\mathrm{el} $ depending on boundary conditions and the `mean field' elastic deformation $\mathbfit{\epsilon}^\mathrm{el}$ due to inhomogeneous distributions of dislocations and a defect related density $\Psibf^\mathrm{def}$, cf. \citet{groma_vi15, kooiman_hg15}. The elastic energy $ \Psibf^\mathrm{el} $ will be discussed in Section \ref{sec: thermodynamic consistency} in the more general setting of curved dislocations which comprises the straight dislocation case as special case. For the defect related energy $\Psibf^\mathrm{def}$ we adopt the form which was derived in \citet{groma_gk07} for the quasi 2D case mostly from scaling arguments. With respect to the functional form, the free energy obtained by \citet{kooiman_hg15} from direct averaging is not much different. We concentrate on the low GND approximation ($| \kappa | << \rho$) for which the defect free energy according to \citet{groma_gk07} assumes the form 
\begin{equation}
	\Psibf^\mathrm{def} = \int{\Psi(\rho,\kappa)} \dex V = \int{ G b^2 \left[ A \rho \ln \left( \frac{\rho}{\rho_0} \right) + B \frac{\kappa^2}{2 \rho} \right] \dex V},
\label{eq: functional 2D}
\end{equation}
with dimensionless constants $A$ and $B$ and a reference dislocation density $ \rho_0 $. \citet{groma_gk07} do not specify the reference density because the evolution of the system remains independent of it.

The back-stress can be derived from the variational derivative of the defect free energy with respect to the net-dislocation density $ \kappa $ as
\begin{equation}
	  \tau_\mathrm{b} = \frac{1}{b} \partial_1 \frac{ \partialvar \Psi(\rho,\kappa)}{\partialvar \kappa} = B G b \partial_1 \left( \frac{ \kappa}{\rho} \right) \approx \frac{B G b}{\rho} \partial_1 \kappa,
\label{eq: taub groma}
\end{equation}
where the latter approximation follows if additionally to $| \kappa | << \rho$ one assume that the gradients $ \partial_1 \rho$ and $ \partial_1 \kappa$ are of the same order of magnitude. As will be discussed in Section \ref{sec: thermodynamic consistency}, the mean field shear stress $ \tau $ derives from the elastic part of the free energy. Consequently, the net shear stress $ \tau_\mathrm{net} =  \tau - \tau_\mathrm{b} $ may be obtained from the above mentioned free energy functional such that (\ref{eq: v with mobility}) may be interpreted as a mobility times a thermodynamics driving force as is common in non-equilibrium thermodynamics. However, the driving force appears to be incomplete in the sense that it does not involve a dependency on the variational derivative of the free energy density with respect to the total dislocation density $ \rho $. That such a term should not appear has been argued for in \citet{groma_gk07}. But this was related to an argument put forth in \citet{groma_cz03}, where a similar term was rationalized away on the ground that dislocations of opposite character when subject to the same stress are bound to move with the same velocity in opposite directions. As we will see below, the additional term derived in the current paper does not rely on allowing for asymmetric dislocation velocities. It will turn out that the negligence of this term is not thermodynamically consistent unless the free energy density would either be independent of $ \rho $ or depending on $\rho$ linearly.
\section{Continuum Dislocation Dynamics for curved dislocations} \label{sec: curved CDD}
Continuum dislocation dynamics comprises a whole hierarchy of kinematic theories obtainable by truncating a series of alignment tensors \citep{hochrainer15} approximating a higher dimensional description \citep{hochrainer_zg07}. In principle CDD is neither restricted to a single slip system nor to purely conservative dislocation motion. However, the only worked out examples so far are the two lowest order closures for single slip systems subject to the assumption of an isotropic dislocation velocity (i.e., independent of dislocation character). In the sequel we will restrict attention to the lowest order truncation with internal variables: total dislocation density $ \rho $, dislocation density vector $ \rhobf $, and dislocation curvature density $ q$. We note that the transfer of the thermodynamic approach presented in the following to higher order truncations and non-isotropic velocities is straight forward.
\subsection{Kinematics of single slip CDD}
We employ the evolution equations in the conservative form for $ q $ as derived in \citet{hochrainer15}. The evolution of the plastic distortions and the dislocation variables then assume the form
\begin{eqnarray} 
	\partial_t \beta^\mathrm{pl}_{ij}  &=&  \rho v b \; n_i m_j,  \label{Eq: evol betap} \\
	\partial_t \rho  &=&  \partial_i \left( v \varepsilon_{ij} \rho_j \right) + v q, \label{Eq: evol rho iso v} \\
	\partial_t \rho_i	&=& -\varepsilon_{i j} \partial_j \left( v \rho \right),  \label{Eq: evol kappa iso v} \\
	\partial_t q &=&  \partial_i \left( v  q_i - \rho_{ji} \partial_j v \right), \label{Eq: evol q iso v}
\end{eqnarray}
where $\varepsilon_{i j} = \upvarepsilon_{ikj} n_k $ denotes the operator performing a cross product with the slip plane normal $ \n $. Again, $v$ denotes a dislocation velocity which is in the current framework assumed to be described by a scalar such that dislocation segments of different character will move with the same speed in directions perpendicular to their line-direction within the glide plane. The curvature vector $q_i$ and the second order dislocation density tensor $ \rho_{ij} $ appearing in the evolution equation of $ q $ (\ref{Eq: evol q iso v}) are unknown and subject to closure assumptions. In the current work we do not specify the closure assumptions and refer to \citet{monavari_zs14} and \citet{hochrainer15} for discussions of this purely kinematic closure. We note that an impressive demonstration of the kinematic consistency of equations (\ref{Eq: evol betap})--(\ref{Eq: evol q iso v}) was recently presented by \citet{sandfeld_p15} in a comparison with simplified discrete dislocation simulations.

The plastic distortions from a single slip system are again characterized by the accumulated plastic slip $ \gamma $, such that $\mathbfit{\beta}^\mathrm{pl} = \gamma \n \otimes \m $. The role of the GND density $ \kappa $ in the quasi 2D theory is now adopted by the dislocation density (or GND) vector which relates to the accumulated plastic slip through
\begin{equation}
	\rho_i = \upvarepsilon_{ijk} \partial_j \frac{\gamma}{b}  n_k = - \frac{1}{b} \varepsilon_{ij} \partial_j \gamma.
\label{eq: rho_i gamma}
\end{equation}

We note that the kinematic evolution equations for straight parallel edge dislocations (\ref{Eq: evol gamma})--(\ref{Eq: evol kappa istvan}) arise as special case of the above evolution equations (\ref{Eq: evol betap})--(\ref{Eq: evol q iso v}) if all quantities are constant in edge direction and the curvature density $q$ vanishes identically. With the coordinate system introduced in Section \ref{sec: constitutive straight} the GND density $ \kappa $ is in this case the 2-component (i.e. edge component) of the dislocation density vector, $ \kappa = \rho_2 $.
\subsection{Thermodynamically consistent single slip CDD} \label{sec: thermodynamic consistency}
While there is currently no free energy available for CDD of curved dislocations it seems natural that a free energy $ \Psibf $ of the CDD system will also be composed of an elastic energy $ \Psibf^\mathrm{el} $, depending on the elastic strain $\mathbfit{\epsilon}^\mathrm{el}$ due to boundary conditions and the `mean field' deformation due to inhomogeneous distributions of dislocations, and a defect related energy $\Psibf^\mathrm{def}$. Both energy contributions arise from integrals over according energy densities which we define through respective terms in the following equation
\begin{equation}
	\Psibf = \Psibf^\mathrm{el} + \Psibf^\mathrm{def} = \int{\frac{1}{2} \mathbfit{\epsilon}^\mathrm{el} : \mathbf{C} : \mathbfit{\epsilon}^\mathrm{el} \dex V} + \int{\Psi(\rho,\rhobf, q)} \dex V.
\label{eq: functional}
\end{equation}

In the sequel we derive a condition for the functional form of the dislocation velocity as a function of the thermodynamic forces defined through the variational derivatives of the free-energy functional. The development of CDD mainly targets at face centered cubic (fcc) crystals, where --aside from high temperature applications-- thermal effects are usually assumed to be negligible \citep{groma_gk07}. We therefore also ignore temperature effects and consequently find the requirement of thermodynamic consistency such that the spontaneous evolution of the system may not increase the free energy. In other words, we require that
\begin{equation}
	\frac{\dex \Psibf}{\dex t} \leq 0.
\label{eq: thermodynamic consistency}
\end{equation}
Assuming that the free energy does not depend explicitly on time the total time derivative is obtained from the chain rule as
\begin{equation}
	\frac{\dex \Psibf}{\dex t} = \int \left(  \frac{\partial \mathbfit{\epsilon}^\mathrm{el}}{\partial t} : \mathbfit{\sigma} + \frac{\partialvar  \Psi}{\partialvar \rho} \frac{\partial \rho}{\partial t} + \frac{\partialvar  \Psi}{\partialvar \rho_i} \frac{\partial \rho_i}{\partial t} + \frac{\partialvar  \Psi}{\partialvar q} \frac{\partial q}{\partial t} \right) \dex V.  
\label{eq: total time derivative}
\end{equation}
For the elastic part we find using partial integration (\ref{eq: partial integration}) and (\ref{Eq: evol betap})
\begin{eqnarray}
	\frac{\dex \Psibf^\mathrm{el}}{\dex t} &=& \int \frac{\partial \epsilon^\mathrm{el}_{ij}} {\partial t} \sigma_{ij} \dex V \nonumber \\
	&=& \int{\left( \partial_i \partial_t u_j - \partial_t \beta^\mathrm{pl}_{ij} \right) \sigma_{ij} \dex V} \nonumber \\
	&=& \int{- \partial_t u_j \partial_i \sigma_{ij} - \rho v b \sigma_{ij} n_i m_j \dex V} \nonumber \\
	&=& \int{- \rho v b \tau \dex V} \label{PK driving force}.
\label{eq:}
\end{eqnarray}
The last identity follows from the solenoidality of the stress tensor (\ref{eq: divergence stress}) and we introduced the resolved (mean-field) shear stress $ \tau = \sigma_{ij} n_i m_j $.

For the explicitly defect related part (note that the mean field stress depends on the GND density via the incompatibility) we find from (\ref{eq: total time derivative}) and repeated application of partial integration
\begin{eqnarray}
	\frac{\dex \Psibf^\mathrm{def}}{\dex t} 	&=& \int \left\{ \frac{\partialvar  \Psi}{\partialvar \rho} \left[ \partial_i \left( v \varepsilon_{ij} \rho_j \right) + v q \right] - \frac{\partialvar  \Psi}{\partialvar \rho_i} \left[ \varepsilon_{i j} \partial_j \left( v \rho \right) \right] + \frac{\partialvar  \Psi}{\partialvar q} \left[ \partial_i \left( v  q_i - \rho_{ji} \partial_j v \right) \right] \right\} \dex V \nonumber \\
%	&=& \int \left[ - \partial_i \frac{\partialvar  \Psi}{\partialvar \rho}  v \varepsilon_{ij} \rho_j + \frac{\partialvar  \Psi}{\partialvar \rho} v q - \partial_i \frac{\partialvar  \Psi}{\partialvar q} \left( v  q_i - \rho_{ji} \partial_j v \right) +  \partial_j \left( \varepsilon_{i j} \frac{\partialvar  \Psi}{\partialvar \rho_i}\right) v \rho \right] \dex V \nonumber \\
%	&=& \int \left[ - v \varepsilon_{ij} \rho_j \partial_i \frac{\partialvar  \Psi}{\partialvar \rho} + v q \frac{\partialvar  \Psi}{\partialvar \rho} - v q_i \partial_i \frac{\partialvar  \Psi}{\partialvar q}  - v \partial_j \left( \rho_{ji} \partial_i \frac{\partialvar  \Psi}{\partialvar q} \right) + v \rho \partial_j \left( \varepsilon_{i j} \frac{\partialvar  \Psi}{\partialvar \rho_i}\right) \right] \dex V \nonumber \\
	&=& \int v \left[ - \varepsilon_{ij} \rho_j \partial_i \frac{\partialvar  \Psi}{\partialvar \rho} + q \frac{\partialvar  \Psi}{\partialvar \rho} + \rho \partial_j \left( \varepsilon_{i j} \frac{\partialvar  \Psi}{\partialvar \rho_i}\right) - q_i \partial_i \frac{\partialvar  \Psi}{\partialvar q}  - \partial_j \left( \rho_{ji} \partial_i \frac{\partialvar  \Psi}{\partialvar q} \right)  \right] \dex V \label{eq: thermodynamic consistency psidef}
\end{eqnarray}
Inserting (\ref{eq: thermodynamic consistency psidef}) and (\ref{PK driving force}) into (\ref{eq: total time derivative}) the requirement for thermodynamic consistency reads
\begin{equation}
	\int{v \left[- \rho b \tau - \varepsilon_{ij} \rho_j \partial_i \frac{\partialvar  \Psi}{\partialvar \rho} + q \frac{\partialvar  \Psi}{\partialvar \rho} + \rho \partial_j \left( \varepsilon_{i j} \frac{\partialvar  \Psi}{\partialvar \rho_i}\right) - q_i \partial_i \frac{\partialvar  \Psi}{\partialvar q}  - \partial_j \left( \rho_{ji} \partial_i \frac{\partialvar  \Psi}{\partialvar q} \right) \right] \dex V} \leq 0.
\label{eq: thermodynamic consistency evolution}
\end{equation}
A thermodynamically consistent dislocation velocity must therefore have the opposite sign than the term in square brackets. This is ensured if the velocity contains the negative multiple of the term in brackets as a factor. We assume the velocity has the same form as in the quasi 2D case,
\begin{equation}
	v = M(\tau_\mathrm{net}, \rho) b \tau_\mathrm{net}. 
\label{eq: consistent v}
\end{equation}
Moreover, we may safely assume that the flow stress in the case of curved dislocations also obeys the Taylor relation $ \tau_\mathrm{f} = \alpha Gb \sqrt{\rho} $ such that the nonlinear mobility function $ M\left( \tau_\mathrm{net} , \rho \right) $ may again be defined by (\ref{eq: mobility}).

With the velocity of the form (\ref{eq: consistent v}), thermodynamic consistency is ensured if the net shear stress $ \tau_\mathrm{net} $ is a negative multiple of the term in brackets in (\ref{eq: thermodynamic consistency evolution}). In order to obtain terms which have the dimension of stress we divide the term in brackets by $ \rho b $ and arrive at
\begin{equation}
	\tau_\mathrm{net} = \tau + \frac{\varepsilon_{ij} \rho_j}{\rho b} \partial_i \frac{\partialvar  \Psi}{\partialvar \rho} - \frac{q}{\rho b} \frac{\partialvar  \Psi}{\partialvar \rho} - \frac{1}{b} \partial_j \left( \varepsilon_{i j} \frac{\partialvar  \Psi}{\partialvar \rho_i} \right) + \frac{q_i}{\rho b} \partial_i \frac{\partialvar  \Psi}{\partialvar q} + \frac{1}{\rho b} \partial_j \left( \rho_{ji} \partial_i \frac{\partialvar  \Psi}{\partialvar q} \right).
\label{eq: taunet curved}
\end{equation}
\subsubsection{Thermodynamic consistency from a postulated energy functional}
In order to understand the appearing mesoscopic stress contributions  in (\ref{eq: taunet curved}) and to relate them to the terms obtained in the straight dislocation case we now \emph{assume} the free energy density in the curved dislocation case in analogous form to the one for straight dislocations as
\begin{equation}
	\Psi(\rho,q,\rhobf) = Gb^2 \left( A \rho \ln \left(\frac{\rho}{\rho_0} \right) + \frac{1}{2} \frac{B_{ij} \rho_i \rho_j}{\rho} + C \frac{1}{2} \frac{q^2}{\rho^2} \right),
\label{eq: quadratic energy}
\end{equation}
with dimensionless constants $ A $, $B_{ij}$ and $ C$ and a reference density $ \rho_0 $. In a spatially and elastically isotropic theory as developed here the tensor $ \mathbfit{B} $ is expected to be spherical, i.e $ B_{ij} = B \updelta_{ij}$, with a constant $ B $. Assuming this form we find
\begin{eqnarray}
	\frac{\partialvar  \Psi}{\partialvar \rho} &=& Gb^2 \left[ A \left(\ln \left(\frac{\rho}{\rho_0}\right) + 1 \right) - \frac{B}{2} \frac{ \rho_i \rho_i}{\rho^2} - C \frac{q^2}{\rho^3}\right]  \label{eq: d psi d rho} \\
	\frac{\partialvar  \Psi}{\partialvar \rho_i} &=& Gb^2 B \frac{\rho_i}{\rho} \\
	\frac{\partialvar  \Psi}{\partialvar q} &=& Gb^2 C \frac{q}{\rho^2}.
\end{eqnarray}
Before we proceed with inserting these expressions into the definition of the net shear stress (\ref{eq: taunet curved}), we make some simplifying assumptions. Please note that these approximations are mostly done here to arrive at readable and partly familiar expressions. In principle any of the following negligences may violate thermodynamic consistency in special situations. As in the quasi 2D theory we assume small GND densities, $ \sqrt{\rho_i \rho_i} << \rho $, and additionally we assume that the average radius of curvature $ r = \rho / q $ is much larger than the average dislocation spacing, i.e. $ r >> 1 / \sqrt{\rho}$. Because the curvature density is $ q = \rho /r $ this implies $ q^2 << \rho^3 $, such that we approximate (\ref{eq: d psi d rho}) by
\begin{equation}
	\frac{\partialvar  \Psi}{\partialvar \rho} \approx Gb^2 A \left(\ln \left(\frac{\rho}{\rho_0} \right) + 1 \right).
\label{eq: approx d psi d rho}
\end{equation}
With regard to the spatial derivatives of the variational derivatives in (\ref{eq: taunet curved}) we again assume that the spatial derivatives of all density variables $ \rho, \rhobf, $ and $ q $ are of the same order of magnitude. We then obtain the approximate expressions
\begin{eqnarray}
	\partial_i \frac{\partialvar  \Psi}{\partialvar \rho} &\approx& A Gb^2 \frac{\partial_i \rho}{\rho}, \\
	\partial_i \frac{\partialvar  \Psi}{\partialvar \rho_j} &\approx& B Gb^2 \frac{\partial_i \rho_j}{\rho}, \\
	\partial_i  \frac{\partialvar  \Psi}{\partialvar q} &\approx& C Gb^2 \frac{\partial_i q}{\rho^2}.
\end{eqnarray}
Inserting these approximate expression into the net shear stress (\ref{eq: taunet curved}) we obtain
\begin{equation}
	\tau_\mathrm{net} \approx \tau + Gb \left\{ \frac{A \varepsilon_{ij} \rho_j}{\rho^2 } \partial_i \rho - \frac{A q}{\rho} \left(\ln \left(\frac{\rho}{\rho_0}\right) + 1 \right) - \frac{B \varepsilon_{i j}}{\rho} \partial_j \rho_i  + \frac{C q_i}{\rho^3 } \partial_i q + \frac{C}{\rho } \partial_j \left( \frac{\rho_{ji}}{\rho^2}  \partial_i q \right)  \right\}.
\label{eq: taunet curved approx}
\end{equation}
We interpret each of the summands in the curly braces (upon multiplication with $Gb$) as a mesoscopic shear stress contribution. We begin with the term which generalizes the back stress found in the theory of straight parallel edge dislocations, which is the third term in the braces. That this is the back stress seems obvious because the term involves a gradient of the GND density. Because the GND density is itself a gradient of the plastic slip, this term is again a second order strain gradient and we find
\begin{equation}
	\tau_\mathrm{b} = \frac{B Gb}{ \rho} \partial_j \left( \varepsilon_{i j} \rho_i \right) = - \frac{B Gb}{\rho} \partial_j \left( \varepsilon_{i j} \varepsilon_{i k} \partial_k \frac{\gamma}{b} \right) = - \frac{B G}{\rho} \left( \updelta_{jk} - n_j n_k \right) \partial_j \partial_k \gamma.
\label{eq: backstress}
\end{equation}
The last expression in (\ref{eq: backstress}) contains the Laplace operator within the glide plane $ \left( \updelta_{ij} - n_i n_j \right) \partial_i \partial_j $. The according identity in (\ref{eq: backstress}) is obtained from the antisymmetry $ \varepsilon_{ji} = -\varepsilon_{ij} $ and the fact that $ \varepsilon_{ik}\varepsilon_{kj}  = - \left( \updelta_{ij} - n_i n_j \right) $. The form of the back stress term is remarkable not only because it naturally generalizes the expression obtained in the quasi 2D theory, but because the Laplace operator of the plastic shear has been postulated in phenomenological strain gradient models used to account for strain gradient effects in small scale plasticity \citep{aifantis87}. Notably, however, the length scale in front of the current back stress expression is consistently found as the only characteristic length scale in a dislocation system, which is the average dislocation spacing governed by the dislocation density.

A further remarkable term is the first term in the curly braces in (\ref{eq: taunet curved approx}), which we call the (density) gradient stress
\begin{equation}
	\tau_\mathrm{g} = \frac{A Gb \varepsilon_{ij} \rho_j}{\rho^2 } \partial_i \rho.
\label{eq: couple}
\end{equation}
This shear stress contribution is of special interest for two reasons: (\emph{i}) it contains the gradient of the total dislocation density and (\emph{ii}) the term stems from the flux part of the total dislocation density evolution and should therefore also appear in a 2D theory of straight parallel edge dislocations. For realizing the latter point we note that the term is a contraction of the density vector $ \rhobf $ tilted by $90^\circ$ in the glide plane and the gradient of $ \rho $. This term therefore reads 
\begin{equation}
	\tau_\mathrm{g}^{2\mathrm{D}} = - \frac{A Gb}{\rho^2} \kappa \partial_1 \rho
\label{eq: couple 2D}
\end{equation}
in the straight dislocation case, where we remind that $ \kappa = \rho_2 $. That the term contains a gradient of the total dislocation density at first sight seems unphysical, as it appears to drive a diffusive behavior for the total dislocation density. However, the gradient stress additionally requires the existence of GNDs, which on the one hand suggests the appearance of a local stress field and on the other hand renders the expression directional such that the contribution is not actually diffusional.

The remaining mesoscopic stress contributions are unique to the theory of curved dislocations. One of them has been postulated earlier by \citet{hochrainer_etal14} as a line tension term. To see the connection we introduce the average dislocation curvature (the inverse of the average radius $ r $ discussed before) which is defined as $ k = q / \rho $. The line tension term has been postulated to be of the form $ \tau_\mathrm{lt} = Tk $ with a line tension $ T \propto Gb $. The according term in (\ref{eq: taunet curved approx}) is the second one in the curly braces which reads
\begin{equation}
	\tau_\mathrm{lt} = A Gb \left(\ln \left(\frac{\rho}{\rho_0}\right) + 1 \right) k.
\label{eq: line tension}
\end{equation}
The current form of the term is irritating because it explicitly depends on the reference density $ \rho_0 $ which is not supposed to appear in the evolution equations according to \citet{groma_gk07}. We shall postpone the discussion of this point to Section \ref{sec: discussion}, and only note at this point, that regardless of the specific form of the free energy, a line-tension type shear stress will appear in thermodynamically consistent constitutive laws unless the free energy would be independent of $ \rho $.

The two stress contributions which contain derivatives of the curvature density, 
\begin{equation}
	\tau_\mathrm{q1} = \frac{C Gb q_i}{\rho^3 } \partial_i q,  \quad \text{and} \quad \tau_\mathrm{q2} = \frac{C Gb}{\rho } \partial_j \left( \frac{\rho_{ji}}{\rho^2}  \partial_i q \right),
\label{eq: q-stresses}
\end{equation}
are completely new and unique to the theory of curved dislocations. As yet we cannot offer physical interpretation of these terms. But we observe that $ \tau_\mathrm{q1} $ is structurally similar to $ \tau_\mathrm{g} $. Note that $ \tau_\mathrm{q1} $ is a contraction of the curvature vector $\q$ and the gradient of the scalar curvature density $q$ similar to $ \tau_\mathrm{g} $ being a contraction of the dislocation density vector $ \rhobf $ and the gradient of the density $ \rho $. The second stress term can be further transformed by employing the fact that the dislocation curvature vector arises as the divergence of the second order alignment tensor, i.e $ q_i = \partial_j \rho_{ji} $ \citep{hochrainer15}. Accordingly, we obtain
\begin{equation}
	\tau_\mathrm{q2} = \tau_\mathrm{q1} + \frac{C Gb \rho_{ji}}{\rho } \partial_j \left( \frac{\partial_i q}{\rho^2}  \right).
\label{eq: tau_q2}
\end{equation}
Because the last term loosely resembles the back stress (in being a second order derivative) we denote it as
\begin{equation}
	\tau_\mathrm{qb} = \frac{C Gb \rho_{ji}}{\rho } \partial_j \left( \frac{\partial_i q}{\rho^2} \right).
\label{eq: tau_qb}
\end{equation}
For harmonizing the appearance we define the gradient curvature stress as $ \tau_\mathrm{qg} := 2 \tau_\mathrm{q1} $ and finally obtain the net shear stress as
\begin{equation}
	\tau_\mathrm{net} = \tau + \tau_\mathrm{g} -  \tau_\mathrm{lt} - \tau_\mathrm{b} + \tau_\mathrm{qg} + \tau_\mathrm{qb}.
\label{eq: consistent taunet}
\end{equation}

We derived the interpretations of the stress contributions for the free energy density (\ref{eq: quadratic energy}) postulated in analogy to the free energy expression presented by \citet{groma_gk07} in the quasi 2D case. But we may transfer these interpretations to any free-energy density, such that we define in general
\begin{eqnarray}
	\tau_\mathrm{g} &:=& \frac{\varepsilon_{ij} \rho_j}{\rho b} \partial_i \frac{\partialvar  \Psi}{\partialvar \rho}, \\ \tau_\mathrm{lt} &:=& \frac{k}{b} \frac{\partialvar  \Psi}{\partialvar \rho}, \\ \tau_\mathrm{b} &:=& \frac{1}{b} \partial_j \left( \varepsilon_{i j} \frac{\partialvar  \Psi}{\partialvar \rho_i} \right), \\  \tau_\mathrm{qg} &:=& \frac{2 q_i}{\rho b} \partial_i \frac{\partialvar  \Psi}{\partialvar q}, \\ \tau_\mathrm{qb} &:=& \frac{\rho_{ji} }{\rho b} \partial_j \left( \partial_i \frac{\partialvar  \Psi}{\partialvar q} \right).
\label{eq: stress contributions general}
\end{eqnarray}
With these definitions entered into the net shear stress (\ref{eq: consistent taunet}) the velocity may also in general be defined by (\ref{eq: consistent v}).

We finally find that a thermodynamic consistent dislocation velocity of the form (\ref{eq: v with mobility}) for the quasi 2D case requires the net shear stress to be of the form
\begin{equation}
	\tau_\mathrm{net}^{2\mathrm{D}} = \tau + \tau_\mathrm{g}^{2\mathrm{D}} - \tau_\mathrm{b}^{2\mathrm{D}},
\label{eq: consistent taunet 2d}
\end{equation}
with
\begin{eqnarray}
	\tau_\mathrm{g}^{2\mathrm{D}} &:=& - \frac{\kappa}{\rho b} \partial_1 \frac{\partialvar  \Psi}{\partialvar \rho}, \\ \tau_\mathrm{b}^{2\mathrm{D}} &:=& \frac{1}{b} \partial_1 \frac{\partialvar  \Psi}{\partialvar \kappa}.
\label{eq: stress contributions general 2D}
\end{eqnarray}
Taking the quasi 2D free energy from (\ref{eq: functional 2D}) the above definitions, in conjunction with the simplifying assumptions employed throughout the current Section, reproduce (\ref{eq: taub groma}) for $ \tau_\mathrm{b}^{2\mathrm{D}} $ and (\ref{eq: couple 2D}) for $ \tau_\mathrm{g}^{2\mathrm{D}} $. Notably, we conclude that the statistical mechanics theory developed by \citet{groma_cz03} is not thermodynamically consistent with the free energy proposed in \citet{groma_gk07}. But for the small GND case treated in the named papers we would expect that the gradient stress (\ref{eq: couple 2D}) would be much smaller than the back stress (\ref{eq: taub groma}), at least when the constants $ A $ and $ B $ in the free energy (\ref{eq: functional 2D}) are of the same order of magnitude.

\section{Discussion} \label{sec: discussion}

In the current paper we present the first thermodynamically consistent phase field-like framework for CDD comprising the variants for straight parallel edge dislocations and the single slip theory based on the internal state variables: total dislocation density $ \rho $, dislocation density vector $ \rhobf $, and dislocation curvature density $ q $. The constitutive equations for the dislocation velocity in CDD had earlier been assumed in analogy to the quasi 2D case which made them appear somewhat ad hoc because they were neither obtained from statistical averaging nor from a general principle in the spirit of non-equilibrium thermodynamics. In the current contribution we showed how this deficiency may be overcome, once a free energy is formulated in terms of the internal state variables of CDD. The results obtained for a free energy functional postulated in analogy to a free energy functional available for straight parallel edge dislocations are quite satisfactory because they justify the form of a back stress term and the appearance of a line tension term, both of which have been postulated in similar form for CDD before \citep{zaiser_etal07,hochrainer_etal14}. Surprisingly, thermodynamic consistency requires the appearance of a so-called gradient stress which is missing in the theory of Groma and co-workers both in the statistical mechanics as in the phase field approach to quasi 2D dislocation systems. With the current derivation we not only find that this term is missing in the straight dislocation case but also derived the vectorial generalization of its definition for CDD of curved dislocations. Besides the line tension term, two additional shear stress contributions were obtained which relate to the curvature of dislocations. These terms are as yet not well understood and it seems too early to analyze them in detail before more is known about the actual dependence of the free energy on the curvature density $ q $. %From what is known about single dislocations, the contribution of the average curvature to the free energy would be expected to be rather small.

The line tension term $ \tau_\mathrm{lt} $ in (\ref{eq: line tension}) was found to be depending on the reference density $ \rho_0 $ used in the logarithmic expression in the energy density. While \citet{groma_gk07} would consider this as contradicting a scale invariant theory, we note that an energy contribution of the form $ \rho \ln \left( \rho / \rho_0 \right) $ has been derived earlier, e.g. by \citet{wilkens69} and \citet{berdichevsky06} which both assign a physical meaning to the reference density $ \rho_0 $. In \citet{wilkens69} the energy of a system of dislocations with density $ \rho $ is reported to have the form $ \rho \ln \left( R_\mathrm{e} / r_0 \right) $, with an outer cut-off radius $ R_\mathrm{e} $ and an inner cut-off radius $ r_0 $, in analogy to the energy of single dislocations \citep{hirth_l68}. For certain random distributions of dislocations the outer cut-off radius is found to scale like the average dislocation spacing, $ R_\mathrm{e} \propto 1 / \sqrt{\rho} $, due to screening effects. The inner cut-off radius is taken as the dislocation core radius which is on the order of the magnitude of the Burgers vector $r_0 \approx b$. We may therefore transform the energy expression of \citet{wilkens69} such that
\begin{equation}
	\rho \ln \left( R_\mathrm{e} / r_0 \right) = A \rho \ln \left( \rho / b^{-2} \right),
\label{eq: energy wilkens}
\end{equation}
with a (negative) constant $A$, which contains all unknown factors. The reference density in this case is $ \rho_0 = b^{-2} $, which is well defined and an obvious upper bound or saturation value for admissible dislocation densities. That $ \rho_0 $ should be an upper bound for the dislocation density was also put forward by \citet{berdichevsky06} who suggested the same functional form for the energetic contribution of the total dislocation density. \citet{berdichevsky06} interprets $ \rho_0 $ as a saturation density of the crystal, and obtains for Al and Ni the value $ \rho_0 \approx 3 \cdot 10 ^{14} \, \mathrm{m}^{-2} $. We shall not dwell upon the question of the correct definition of $\rho_0$ at this point. The important message to take away is that the reference density is not an arbitrary scaling parameter which has to drop out of the evolution equations, but a well defined physical quantity. As was pointed out to me by I. Groma, the quasi 2D system is scale free due to the $ 1/r $-dependence of the dislocation interactions, while in 3D systems dislocation core effects as the line tension may well introduce a further parameter with length dimension, as e.g., the core radius. This line of thought is in accordance with the above cited work by Wilkens.
%\begin{equation}
%	\rho \ln \left( \sqrt{ \rho^{-1} / b^2 } \right) = 0.5 \rho \ln \left[ \left( \rho / b^{-2} \right)^{-1} \right] = - 0.5 \rho \ln \left( \rho / b^{-2} \right) 
%\label{eq:}
%\end{equation}

As the combination of flow stress and net shear stress in CDD essentially translates into the flow rule for phenomenological plasticity laws it is also remarkable that the back stress expression obtained in the current paper resembles a second order strain gradient expression postulated by \citet{aifantis87} in a phenomenological theory before. An important difference to the latter theory is, however, that the prefactor, which is required to have the dimension of a length squared, is not a constant but $ 1 / \rho $, that is the square of the average dislocation spacing $1/ \sqrt{\rho}$, which is the only characteristic length scale of the system. The newly found gradient stress additionally yields a dependence of the `flow rule' on the first order strain gradient, i.e. the GND tensor. While such dependencies have been postulated numerous times in strain gradient plasticity since the pioneering works of \citet{ashby70, fleck_h93, nix_g98, gao_etal99} to name just a few, GNDs were usually assumed to increase the flow stress. By contrast, the gradient stress derived in the current work is a signed quantity, which may obstruct or facilitate plastic deformation depending on the relative direction of the gradient of plastic slip and the gradient of total dislocation density. Furthermore, the required coefficient is again not a constant but involves the total dislocation density and its gradient.

The herein derived theory of course remains preliminary in many respects. Most obviously, the free energy functional is not yet known for the considered single slip CDD theory. But this problem is common to all available strain gradient plasticity theories, as well. Note, however, that CDD is solely based on physically defined internal variables such that there is reason to expect that an according free energy density \emph{is} derivable from dislocation theory . In fact, during the review process Michael Zaiser proposed a framework for deriving the free energy for systems of curved dislocations in terms of the dislocation density vector and the second order dislocation alignment tensor using pair correlation functions \citep{zaiser15}. Another open problem is the form of the mobility function which remains on the level of an educated guess so far. As for the flow stress we note that the Taylor relation is a universal finding at least in fcc crystals. Because the dimensionless prefactor $ \alpha $ is known to fall into a narrow range, $ 0.2 \leq \alpha \leq 0.5 $ and to be only weakly depending on details of the dislocation structure it seems feasible to treat it as a fitting parameter rather than trying to obtain it from `first principles'. Of course, the mobility may also be defined to yield a smoother transition to plastic flow around the flow stress.

Another point obviously missing in the current paper is the consideration of boundary conditions. The decision to neglect boundary conditions was mostly done to keep the paper focused. The derivation of boundary conditions is possible through the application of Gauss' integration theorem in place of the partial integration (\ref{eq: partial integration}) in the time evolution of the free energy.

In the current derivation we solely developed a single slip theory. But of course any crystal plasticity theory needs to be designed for multiple active slip systems. The most simple option would be to treat multiple slip mostly as a superposition of single slips -- with the exception that the flow stress needs to consider dislocations on other slip systems $ s $, e.g. in the generalized Taylor relation \citep{franciosi_bz80} $ \tau^s_\mathrm{f} = Gb \sqrt{\sum_{s'} \alpha_{s s'} \rho_{s'} } $, with dimensionless parameters $ \alpha_{s s'} $ characterizing the `frictional interaction strength' between slip systems $ s $ and $s'$. Truly multiple slip theories will of course need to consider shear stress contributions from correlations between dislocations on different slip systems. The derivation of such a theory will require further research on the kinematic as well as on the energetic/kinetic side.

In the present paper we concentrated on the lowest order CDD theory for curved dislocations. In \citet{hochrainer15} it was demonstrated how more refined single slip CDD theories may be systematically derived from an alignment tensor expansion of a higher dimensional dislocation theory. In the very paper two extensions are sketched, which is (i) a theory again employing an isotropic dislocation velocity $ v $ but considering the evolution of the second order dislocation alignment tensor $\rho_{ij}$, and (ii) a theory for anisotropic velocities described by a second order velocity tensor $v_{ij}$. Assuming that the dependency of the free energy on higher order tensors is known, as suggested by the recent work of \citet{zaiser15}, the derivation of a thermodynamically consistent dislocation velocity put forward in this paper can be straight-forwardly applied to these more refined theories. Maybe the most surprising finding of the current paper is that the admissible and always size-dependent kinetic equations are essentially dictated by the kinematic equations of CDD.

\section*{Acknowledgment}

The author gratefully acknowledges funding by the German Science Foundation DFG under project HO 4227/3-1 and the DFG Research Unit `Dislocation based plasticity' FOR 1650 under project HO 4227/5-1. Furthermore, I thank Michael Zaiser and Stefan Sandfeld for discussing and improving preliminary versions of this manuscript.

%%\bibliographystyle{elsarticle-harv}
%%\bibliography{../../../../references_thomas}

\end{document}